\newcounter{myctr}

\documentclass{ws-ijqi}

\begin{document}

\markboth{H.-T. Elze}
{Proliferation of observables and measurement in quantum-classical hybrids}

\catchline{}{}{}{}{}

\title{PROLIFERATION OF OBSERVABLES AND MEASUREMENT IN QUANTUM-CLASSICAL HYBRIDS  
}

\author{HANS-THOMAS ELZE 
}

\address{Dipartimento di Fisica ``Enrico Fermi'', Universit\`a di Pisa, \\ 
Largo Pontecorvo 3, I-56127 Pisa, Italia
\\ elze@df.unipi.it}



\maketitle

\begin{history}
\received{\today}
\end{history}

\begin{abstract}
Following a review of quantum-classical hybrid dynamics, 
we discuss the ensuing proliferation of observables and relate it to measurements of 
(would-be) quantum mechanical degrees of freedom performed by (would-be) classical ones 
(if they were separable). -- Hybrids consist in coupled classical (``CL'') and 
quantum mechanical (``QM'') objects. Numerous consistency requirements for their description 
have been discussed and are fulfilled here. 
We summarize a representation of quantum mechanics in terms of classical analytical 
mechanics which is naturally extended to QM-CL hybrids.  
This framework allows for superposition, separable, and entangled states originating in 
the QM sector, admits experimenter's `Free Will', and is local and 
non-signalling. -- Presently, we study the set of hybrid observables, 
which is larger than the Cartesian product of QM and CL observables of its components; 
yet it is smaller than a corresponding product of all-classical observables. Thus, {\it quantumness and classicality infect each other}.   
\end{abstract}

\keywords{quantum-classical hybrid dynamics; oscillator representation; classical control; measurement; hybrid observables} 


\section{Introduction}

The hypothetical direct coupling of quantum mechanical and classical degrees of 
freedom -- {\it ``hybrid dynamics''} -- departs from quantum mechanics. We review  
the theory presented in Refs.\,\cite{me11,me12}, emphasizing here 
new aspects of relevant sets of observables related to typical 
hybrid interactions, which resemble measurements.  
 
Hybrid dynamics has been researched for practical as well as theoretical reasons. 
For example, the Copenhagen interpretation, as in standard textbooks, entails the  
measurement problem which, together with the fact that quantum mechanics needs  
interpretation, in order to be operationally well defined, may indicate that 
it deserves amendments. It has been recognized earlier that 
a theory which {\it dynamically} crosses the quantum-classical border should have an 
impact on the measurement problem \cite{Sudarshan123} as well as on  
attempts to describe consistently the interaction between quantum matter and classical 
spacetime \cite{BoucherTraschen}.  
   
Numerous works have appeared attempting to formulate a satisfactory hybrid dynamics. 
Generally, they show deficiencies in one or another respect. 
Which has led to various no-go theorems, in view of a lengthy list of desirable properties 
or consistency requirements, see, for example, 
Refs.\,\cite{CaroSalcedo99,DiosiGisinStrunz}:  
\begin{itemize} 
\item Conservation of energy. 
\item Conservation and positivity of probability. 
\item Separability of QM and CL subsystems in the absence of their interaction, 
recovering the correct QM and CL equations of motion, respectively. 
\item Consistent definitions of states and observables; existence of a Lie bracket structure 
on the algebra of observables that suitably generalizes  
Poisson and commutator brackets. 
\item Existence of canonical transformations generated by the observables; 
invariance of the classical sector under canonical transformations 
performed on the quantum sector only and {\it vice versa}. 
\item Existence of generalized Ehrenfest relations ({\it i.e.} the 
correspondence limit) which, 
for bilinearly coupled CL and QM oscillators,  
are to assume the form of the CL equations of motion  
(``Peres-Terno benchmark'' test \cite{PeresTerno}). 
\end{itemize} 
More recently, these have been followed by more sophisticated considerations trying 
to single out ``{\it the}'' hybrid theory. These require: 
\begin{itemize} 
\item `Free Will' \cite{Diosi11}. 
\item Locality. 
\item No-signalling. 
\item QM / CL symmetries and ensuing separability carry over to hybrids. 
\end{itemize} 

These issues have been reviewed in recent works by Hall and Reginatto, who   
introduced a form of hybrid dynamics that conforms with the first group of points 
listed above \cite{HallReginatto05,Hall08,ReginattoHall08}. 
Their ensemble theory is based on configuration space, which requires a certain 
nonlinearity of the underlying action functional and entails 
effects that might allow to falsify this proposal experimentally \cite{HallEtAll11}. 

We have proposed an alternative theory 
of hybrid dynamics based on notions of phase space \cite{me11}. 
This is partly motivated by work on related topics of general linear dynamics and 
classical path integrals \cite{EGV11,EGV10} and extends work by Heslot, demonstrating   
that quantum mechanics can entirely be rephrased in the language and 
formalism of classical analytical mechanics \cite{Heslot85}. 
Introducing unified notions 
of states on phase space, observables, canonical transformations, and a generalized 
quantum-classical Poisson bracket, this has led to an intrinsically 
linear hybrid theory, which allows to fulfil {\it all} of the above consistency requirements. 

It has been shown more recently by Buri\'c and collaborators that dynamical aspects of our 
proposal can indeed be derived for an all-quantum mechanical composite system by 
imposing constraints on fluctuations in one subsystem, followed by suitable coarse-graining 
\cite{Buric12}.  

Objects that somehow reside between classical and quantum mechanics 
have been described also in a statistical theory, based on very different premises 
than the hybrid theories considered here \cite{Zwitters}. It remains to 
uncover their relation. 

We point out that it is of great practical interest to better 
understand QM-CL hybrids appearing in QM approximation 
schemes. These typically address many-body systems or interacting fields, 
which are naturally separable into QM and CL subsystems, for example, representing fast and 
slow degrees of freedom, mean fields and fluctuations, {\it etc.}  
(keywords: Born-Oppenheimer approximation, mesoscopic systems, CL control of QM objects, 
``semiclassical quantum gravity''); for references see Ref.\,\cite{me11}.  
 
Furthermore, concerning the hypothetical emergence of quantum mechanics from some  
coarse-grained deterministic dynamics (see, for example, Refs.\,\cite{tHooft10,Elze09a,Adler} 
with numerous references to earlier work), the quantum-classical backreaction problem might   
appear in new form, namely regarding the interplay of fluctuations among underlying deterministic 
and emergent QM degrees of freedom. Which can be 
rephrased succinctly as the question: {\it ``Can quantum mechanics be seeded?''}

Besides constructing the QM-CL hybrid formalism and showing how 
it conforms with the above consistency requirements, we earlier discussed  
the possibility to have classical-environment induced decoherence, quantum-classical 
backreaction, a deviation from the Hall-Reginatto proposal in presence 
of translation symmetry, and closure of the algebra of hybrid observables \cite{me11}. 
Questions of locality, symmetry vs. separability, incorporation of superposition, separable, 
and entangled QM states, and `Free Will' were considered in detail \cite{me12}. 

Presently, we briefly recollect, in Section~2., some of the earlier results, 
which will be useful in the following. In Section~3., we expand on our 
previous observation that genuine QM-CL hybrids have an algebra of observables, 
which is not simply formed by the Cartesian product of those pertaining to the QM and CL 
sectors, respectively, in the absence of the hybrid interaction \cite{me11,me12}. 
In various ways, this has more recently been mentioned also in Refs.\,\cite{Buric12b,Barcelo12}. 
We will show here that the characteristically enlarged hybrid algebra of observables can 
actually be understood as a measurement effect exerted by (would-be) CL degrees of freedom 
on (would-be) QM ones (if they were separable). Some concluding remarks are made in Section~4.  

\section{Linear quantum-classical hybrid dynamics -- a review}
 
The following is a synopsis of classical Hamiltonian mechanics,  of 
its generalization incorporating quantum mechanics by Heslot \cite{Heslot85},  
and of our extension which describes the hypothetical 
direct coupling between QM and CL degrees of freedom in hybrids \cite{me11,me12}. 
Readers familiar with the earlier derivations may directly pass to Section~3. 

\subsection{Classical mechanics}
 
Evolution of a {\it classical} object is described in relation to its $2n$-dimensional 
phase space, which is its {\it state space}. A real-valued regular 
function on the state space defines an {\it observable}, {\it i.e.}, a differentiable function 
on this smooth manifold. 

There always exist (local) systems of so-called 
{\it canonical coordinates}, commonly denoted by $(x_k,p_k),\; k=1,\dots ,n$, 
such that the {\it Poisson bracket} of any pair of observables $f,g$ assumes 
the standard form (Darboux's theorem) \cite{Arnold}: 
\begin{equation}\label{PoissonBracket} 
\{ f,g \}\; =\; 
\sum_k\Big (\frac{\partial f}{\partial_{x_k}}\frac{\partial g}{\partial_{p_k}}
-\frac{\partial f}{\partial_{p_k}}\frac{\partial g}{\partial_{x_k}}\Big ) 
\;\;. \end{equation} 
This is consistent with $\{ x_k,p_l\}=\delta_{kl}$, $\{ x_k,x_l\} =\{ p_k,p_l\}=0,\; 
k,l=1,\dots ,n$, and reflects the bilinearity, antisymmetry, derivation-like 
product formula, and Jacobi identity which define a Lie bracket operation, 
$f,g\rightarrow\{ f,g\}$, mapping a pair of observables to an observable.  

General transformations ${\cal G}$ of the state space are restricted by compatibility with 
the Poisson bracket structure to so-called {\it canonical transformations}, which   
do not change physical properties of an object; e.g., a translation, 
a rotation, a change of inertial frame, or evolution in time. Such ${\cal G}$ 
induces a change of an observable, $f\rightarrow {\cal G}(f)$, and is an automorphism 
of the state space compatible with the Poisson brackets.
The set of canonical transformations has a Lie group structure. Therefore, it is sufficient 
to consider infinitesimal transformations generated by the elements of the 
corresponding Lie algebra. Then, an {\it infinitesimal transformation} ${\cal G}$ is 
{\it canonical}, if and only if for any observable $f$ the map $f\rightarrow {\cal G}(f)$ 
is given by $f\rightarrow f'=f+\{ f,g\}\delta\alpha$, with some observable $g$, 
the so-called {\it generator} of ${\cal G}$, and $\delta\alpha$ an infinitesimal real 
parameter. -- An infinitesimal canonical transformation of the canonical coordinates, 
for example, is given by: 
\begin{eqnarray}\label{xcan} 
x_k&\rightarrow &x_k'=x_k+\frac{\partial g}{\partial p_k}\delta\alpha 
\;\;, \\ [1ex] \label{pcan}  
p_k&\rightarrow &p_k'=p_k-\frac{\partial g}{\partial x_k}\delta\alpha 
\;\;, \end{eqnarray} 
where we employ the Poisson bracket of Eq.\,(\ref{PoissonBracket}). 

This illustrates the fundamental relation between observables and generators 
of infinitesimal canonical transformations in classical Hamiltonian mechanics. 

\subsection{Quantum mechanics}

Following Heslot's work, we learn that the previous analysis 
can be generalized and applied to quantum mechanics; this concerns 
the dynamical aspects as well as the notions of states, canonical transformations, and 
observables. 

We recall that the Schr\"odinger equation and its adjoint can be 
derived from an action principle as related Hamiltonian equations of motion 
\cite{me11}. -- We must add the {\it normalization condition}, for any state vector $|\Psi\rangle$: 
\begin{equation}\label{normalization} 
{\cal C}:=\langle\Psi (t)|\Psi (t)\rangle\stackrel{!}{=}\mbox{constant}\equiv 1 
\;\;, \end{equation} 
an essential ingredient of the associated probability interpretation. 
Furthermore, state vectors that differ by an {\it unphysical constant phase} are to be 
identified. Which reminds us that the {\it quantum mechanical state space} 
is formed by the rays of an underlying Hilbert space, {\it i.e.}, forming a complex 
projective space. 

\subsubsection{Oscillator representation}

A unitary transformation describes QM evolution, 
$|\Psi (t)\rangle =\hat U(t-t_0)|\Psi (t_0)\rangle$, 
with $U(t-t_0)=\exp [-i\hat H(t-t_0)/\hbar ]$, 
which solves the Schr\"odinger equation. Thus, a stationary state, characterized by 
$\hat H|\phi_i\rangle =E_i|\phi_i\rangle$, with real energy eigenvalue $E_i$,  
performs a harmonic motion, {\it i.e.}, 
$|\psi_i(t)\rangle =\exp [-iE_i(t-t_0)/\hbar ]|\psi_i(t_0)\rangle
\equiv\exp [-iE_i(t-t_0)/\hbar ]|\phi_i\rangle$. We assume a denumerable set of 
eigenstates of the Hamilton operator $\hat H$. 
 
These findings and the Hamiltonian character of the underlying 
equations of motion suggest to introduce the {\it oscillator representation}. 
Consider expanding any state vector with respect to a complete 
orthonormal basis, $\{ |\Phi_i\rangle\}$:  
\begin{equation}\label{oscillexp} 
|\Psi\rangle =\sum_i|\Phi_i\rangle (X_i+iP_i)/\sqrt{2\hbar} 
\;\;, \end{equation} 
where the generally time dependent expansion coefficients are written 
in terms of real and imaginary parts, $X_i,P_i$ \cite{Heslot85}. 
Employing this expansion allows to   
evaluate what we {\it define} as {\it Hamiltonian function}, {\it i.e.},   
${\cal H}:=\langle\Psi |\hat H|\Psi\rangle$: 
\begin{equation}\label{HamiltonianQM1} 
{\cal H}=\frac{1}{2\hbar}\sum_{i,j}\langle\Phi_i|\hat H|\Phi_j\rangle (X_i-iP_i)(X_j+iP_j) 
=:{\cal H}(X_i,P_i) 
\;\;. \end{equation} 
Choosing the set of energy eigenstates, $\{ |\phi_i\rangle\}$, 
as basis of the expansion, we obtain:     
\begin{equation}\label{HamiltonianQM2} 
{\cal H}(X_i,P_i)=\sum_i\frac{E_i}{2\hbar}(P_i^{\;2}+X_i^{\;2}) 
\;\;, \end{equation} 
hence the name {\it oscillator representation}.  

Indeed, evaluating $|\dot\Psi\rangle =
\sum_i|\Phi_i\rangle (\dot X_i+i\dot P_i)/\sqrt{2\hbar}$ 
according to Hamilton's equations with the Hamiltonian function of 
Eq.\,(\ref{HamiltonianQM1}) or (\ref{HamiltonianQM2}), gives back the Schr\"odinger equation. 
Furthermore, the {\it constraint}, Eq.\,(\ref{normalization}), becomes: 
\begin{equation}\label{oscillnormalization} 
{\cal C}(X_i,P_i)=\frac{1}{2\hbar}\sum_i(X_i^{\;2}+P_i^{\;2})\stackrel{!}{=}1 
\;\;. \end{equation} 
Thus, the vector with components given by 
$(X_i,P_i),\; i=1,\dots ,N$, is 
confined to the surface of a $2N$-dimensional sphere with radius $\sqrt{2\hbar}$, 
which presents a major difference to CL Hamiltonian mechanics. 

Our reasoning, so far, indicates that $(X_i,P_i)$ 
may serve as {\it canonical coordinates} for the state space of a 
QM object. Next, we introduce a {\it Poisson bracket}, 
similarly as in Subsection~2.1., 
for any two {\it observables} on the {\it spherically compactified state space}, 
{\it i.e.} real-valued regular functions $F,G$ of the coordinates $(X_i,P_i)$:     
\begin{equation}\label{QMPoissonBracket} 
\{ F,G \}\; =\; 
\sum_i\Big (\frac{\partial F}{\partial_{X_i}}\frac{\partial G}{\partial_{P_i}}
-\frac{\partial F}{\partial_{P_i}}\frac{\partial G}{\partial_{X_i}}\Big ) 
\;\;. \end{equation} 
Thus, time evolution of any observable $O$ is generated by the Hamiltonian: 
\begin{equation}\label{evolution}  
\frac{\mbox{d}O}{\mbox{d}t}=\partial_tO+\{ O,{\cal H} \} 
\;\;, \end{equation} 
and, in particular, we find that the constraint of Eq.\,(\ref{oscillnormalization}) 
is conserved:  
\begin{equation}\label{constraint} 
\frac{\mbox{d}{\cal C}}{\mbox{d}t}=\{ {\cal C},{\cal H} \}=0 
\;\;. \end{equation} 

\subsubsection{Canonical transformations and quantum observables} 

In the following, we explain the compatibility of the notion of 
observable introduced in passing above -- as in classical mechanics -- with the usual 
QM one. 
This can be demonstrated rigourously by the implementation of canonical transformations  
and analysis of the role of observables as their generators.    

The Hamiltonian function has been defined as observable in  
Eq.\,(\ref{HamiltonianQM1}), which relates it directly to the corresponding QM   
observable, namely the expectation of the self-adjoint Hamilton operator. 
This is indicative of the general structure. -- 
Refering to Refs.\,\cite{me11,Heslot85} for more details, three most important points are:  
\\ \noindent $\bullet$ 
A) {\it Compatibility of unitary transformations and Poisson structure.} -- 
The canonical transformations discussed in Section~2.1. represent 
automorphisms of classical state space which are compatible with the Poisson brackets. 
Automorphisms of QM Hilbert space are implemented by 
unitary transformations. This implies a transformation of the canonical 
coordinates here, {\it i.e.}, of the expansion coefficients $X_i,P_i$ introduced in 
Eq.\,(\ref{oscillexp}). From this, one derives the invariance of   
the Poisson brackets under unitary transformations. 
Consequently, {\it unitary transformations on Hilbert space are canonical transformations 
on the $(X,P)$ state space}.   
\\ \noindent $\bullet$ 
B) {\it Self-adjoint operators as observables.} -- 
Any infinitesimal unitary transformation $\hat U$ can be generated by a self-adjoint operator 
$\hat G$, such that: 
\begin{equation}\label{Uinfini} 
\hat U=1-\frac{i}{\hbar}\hat G\delta\alpha 
\;\;, \end{equation} 
which leads to the QM relation between an observable and 
a self-adjoint operator. By a simple calculation, one obtains: 
\begin{eqnarray}\label{Xcan} 
X_i&\rightarrow &X_i'=X_i+\frac{\partial \langle\Psi |\hat G|\Psi\rangle}
{\partial P_i}\delta\alpha 
\;\;, \\ [1ex] \label{Pcan}  
P_i&\rightarrow &P_i'=P_i-\frac{\partial \langle\Psi |\hat G|\Psi\rangle}
{\partial X_i}\delta\alpha 
\;\;. \end{eqnarray} 
Then, the  
relation between an observable $G$, defined in analogy to Section~2.1., and a self-adjoint 
operator $\hat G$ can be inferred from Eqs.\,(\ref{Xcan})--(\ref{Pcan}): 
\begin{equation}\label{goperator} 
G(X_i,P_i)=\langle\Psi |\hat G|\Psi\rangle 
\;\;, \end{equation} 
{\it i.e.}, by comparison with the classical result.  
Hence, a {\it real-valued regular function $G$ of the state is an observable, if 
and only if there exists a self-adjoint operator $\hat G$ such that Eq.\,(\ref{goperator}) 
holds}. This implies that {\it all QM observables are quadratic forms} 
in the $X_i$'s and $P_i$'s, which are essentially fewer than in the corresponding CL 
case; for the generalization necessary when QM-CL hybrids interact, see Section~3.  
\\ \noindent $\bullet$ 
C) {\it Commutators as Poisson brackets.} -- 
From the relation (\ref{goperator}) between observables and self-adjoint operators and the  
Poisson bracket (\ref{QMPoissonBracket}) one derives: 
\begin{equation}\label{QMPBComm} 
\{ F,G\}=\langle\Psi |\frac{1}{i\hbar}[\hat F,\hat G]|\Psi\rangle 
\;\;, \end{equation}  
with both sides of the equality considered as functions of the variables $X_i,P_i$ 
and with the commutator defined as usual. Therefore,  
the {\it commutator is a Poisson bracket with respect to the $(X,P)$ state space} and 
relates the CL algebra of observables, cf. Section~2.1.,  
to the QM algebra of self-adjoint operators.   

In conclusion, quantum mechanics shares with classical mechanics an even dimensional state 
space, a Poisson structure, and a related algebra of observables. It  
differs essentially by a restricted set of observables and the requirements 
of phase invariance and normalization, which compactify the underlying Hilbert space 
to the complex projective space formed by its rays.  

\subsection{Quantum-classical Poisson bracket, hybrid states and their evolution}

The far-reaching parallel of classical and quantum mechanics, as we have seen,  
suggests a {\it generalized Poisson bracket} for any two observables $A,B$ 
defined on the Cartesian product state space of CL {\it and} QM sectors of a hybrid:  
\begin{eqnarray}\label{GenPoissonBracket} 
\{ A,B\}_\times &:=&\{ A,B\}_{\mbox{\scriptsize CL}}+\{ A,B\}_{\mbox{\scriptsize QM}}
\\ [1ex] \label{GenPoissonBracketdef} 
&:=&\sum_k\Big (\frac{\partial A}{\partial_{x_k}}\frac{\partial B}{\partial_{p_k}}
-\frac{\partial A}{\partial_{p_k}}\frac{\partial B}{\partial_{x_k}}\Big )+  
\sum_i\Big (\frac{\partial A}{\partial_{X_i}}\frac{\partial B}{\partial_{P_i}}
-\frac{\partial A}{\partial_{P_i}}\frac{\partial B}{\partial_{X_i}}\Big ) 
\;\;. \end{eqnarray} 
It shares the usual properties of a Poisson bracket. -- Note that due to the 
convention  introduced by Heslot \cite{Heslot85}, to which we have adhered in 
Section~2.2, in particular,  the QM variables $X_i,P_i$ have dimensions of 
(action$)^{1/2}$ and, consequently, no $\hbar$ appears in 
Eqs.\,(\ref{GenPoissonBracket})--(\ref{GenPoissonBracketdef}). At the expense 
of introducing appropriate rescalings, these variables could be made to have their 
usual dimensions and $\hbar$ to appear explicitly here. However, for the remainder  
of this article, 
we choose units conveniently such that $\hbar\equiv 1$.
 
Let an {\it observable ``belong'' to the CL (QM) sector, if it is  
constant with respect to the canonical coordinates of the QM (CL) sector}. Then,   
the $\{\;,\;\}_\times$-bracket has the important properties: 
\\ \noindent $\bullet$ 
D) It reduces to the Poisson brackets introduced 
in Eqs.\,(\ref{PoissonBracket}) and (\ref{QMPoissonBracket}), respectively,   
for pairs of observables that belong {\it either} to the CL {\it or} the QM sector. 
$\bullet$ 
E) It reduces to the appropriate one of the former brackets, 
if one of the observables belongs only to either one of the two sectors. 
$\bullet$ 
F) It reflects the {\it separability} of CL and QM sectors, 
since $\{ A,B\}_\times =0$, if $A$ and $B$ belong to different sectors.  
Hence, {\it if a canonical tranformation 
is performed on the QM (CL) sector only, then observables that belong to the 
CL (QM) sector remain invariant.}

Next, we recall the hybrid density 
$\rho$ defined in Ref.\,\cite{me11} 
as expectation in a given state $|\Psi\rangle$ of a 
self-adjoint density operator $\hat\rho$:  
\begin{equation}\label{rhodens}
\rho (x_k,p_k;X_i,P_i):=\langle\Psi |\hat\rho (x_k,p_k)|\Psi\rangle
=\frac{1}{2}\sum_{i,j}\rho_{ij}(x_k,p_k)(X_i-iP_i)(X_j+iP_j)
\;\;, \end{equation} 
using the oscillator expansion, Eq.\,(\ref{oscillexp}), and   
$\rho_{ij}(x_k,p_k):=\langle\Phi_i|\hat\rho (x_k,p_k)|\Phi_j\rangle 
=\rho_{ji}^\ast (x_k,p_k)$. It describes a {\it quantum-classical hybrid ensemble} 
by a real-valued, positive semi-definite, normalized, and possibly time dependent 
regular function, the probability distribution $\rho$, on the Cartesian 
product state space canonically coordinated by $2(n+N)$-tuples $(x_k,p_k;X_i,P_i)$; 
the variables $x_k,p_k,\; k=1,\dots ,n$ and $X_i,P_i,\; i=1,\dots ,N$ are reserved 
for the CL and QM sector, respectively.

Expanding $\hat\rho$ in terms of its eigenstates,   
$\hat\rho =\sum_jw_j|j\rangle\langle j|$, one also obtains: 
\begin{eqnarray}\label{rhointerpr1}
\rho (x_k,p_k;X_i,P_i)&=&\sum_jw_j(x_k,p_k)\mbox{Tr}(|\Psi\rangle\langle\Psi |j\rangle\langle j|) 
\\ [1ex] \label{rhointerpr2}
&=&\sum_jw_j(x_k,p_k)|\langle j|\Psi\rangle |^2  
\;\;,  \end{eqnarray} 
with $0\leq w_j\leq 1$ and $\sum_j\int\Pi_l(\mbox{d}x_l\mbox{d}p_l)w_j(x_k,p_k)=1$. -- 
This suggests that $\rho (x_k,p_k;X_i,P_i)$, when properly  
normalized, is the {\it probability density to find in the hybrid ensemble the QM state} 
$|\Psi\rangle$, parametrized by $X_i,P_i$ through Eq.\,(\ref{oscillexp}), {\it together with  
the CL state} given by a point in phase space, specified by coordinates $x_k,p_k$.   

The content of our definition of the hybrid density $\rho$ has been further investigated 
in Ref.\,\cite{me12} with respect to superposition, pure/mixed, or separable/entangled 
QM states, possibly present before or after QM and CL sectors of a hybrid interact. 

Instead of pursuing this, we introduce the appropriate {\it Liouville equation} 
for the dynamical evolution of hybrid ensembles \cite{me11}. 
Based on Liouville's theorem and the generalized Poisson bracket defined in 
Eqs.\,(\ref{GenPoissonBracket})--(\ref{GenPoissonBracketdef}), we are led to: 
\begin{equation}\label{rhoevol} 
-\partial_t\rho = \{\rho ,{\cal H}_\Sigma\}_\times  
\;\;, \end{equation}
with ${\cal H}_\Sigma\equiv{\cal H}_\Sigma (x_k,p_k;X_i,P_i)$ and:  
\begin{equation}\label{HtotalInt} 
{\cal H}_\Sigma:={\cal H}_{\mbox{\scriptsize CL}}(x_k,p_k)
+{\cal H}_{\mbox{\scriptsize QM}}(X_i,P_i) 
+{\cal I}(x_k,p_k;X_i,P_i)   
\;\;, \end{equation} 
which defines the relevant {\it Hamiltonian function}, including a hybrid interaction;  
${\cal H}_\Sigma$ is required to be an {\it observable}, in order to have 
a meaningful notion of energy. Note that {\it energy conservation} follows trivially from 
$\{{\cal H}_\Sigma,{\cal H}_\Sigma\}_\times =0$.  

An important advantage of Hamiltonian flow and a general property of the Liouville 
equation in this context is \cite{Arnold}:  
\\ \noindent $\bullet$
G) The normalization and positivity of the probability 
density $\rho$ are conserved in presence of 
a hybrid interaction; hence, its interpretation remains valid. 

However, the simple form of $\rho$ as a bilinear function of QM ``phase space'' 
variables $X_i,P_i$, stemming from the expectation of a density operator $\hat\rho$, 
does not hold generally for interacting QM-CL hybrids. 
As pointed out in Section~5.4 of Ref.\,\cite{me11}, the oscillator 
expansion of observables, such as in the second of Eqs.\,(\ref{rhodens}), 
has to be generalized to allow for what we named   
{\it almost-classical observables} next.   

\section{Proliferation of observables by measurement-like interactions}

This comes about, since the ``classical part'' of the bracket, 
$\{ A,B\}_{\mbox{\scriptsize CL}}$, can generate terms which do {\it not} qualify as 
observable with respect to the QM sector; here we assume that $A$ and 
$B$ are both hybrid observables, as defined before. 
Such terms, in general, are of the form: 
\begin{eqnarray} 
&\;&\frac{1}{4}\sum_{i,i',j,j'}\{A_{ij},B_{i'j'}\}_{\mbox{\scriptsize CL}} 
(X_i-iP_i)(X_j+iP_j)(X_{i'}-iP_{i'})(X_{j'}+iP_{j'}) 
\nonumber \\ [1ex] \label{nonlinear} 
&\;&=\sum_{i,i',j,j'}\langle\Psi |\Phi_i\rangle\langle\Psi |\Phi_{i'}\rangle 
\{A_{ij},B_{i'j'}\}_{\mbox{\scriptsize CL}} 
\langle\Phi_j|\Psi\rangle\langle\Phi_{j'}|\Psi\rangle
\;\;, \end{eqnarray} 
where we used the oscillator expansion, Eq.\,({\ref{oscillexp}), and: 
\begin{equation}\label{nonlinear1}
\{A_{ij},B_{i'j'}\}_{\mbox{\scriptsize CL}} 
=\sum_k
\Big (\frac{\partial A_{ij}}{\partial_{x_k}}\frac{\partial B_{i'j'}}{\partial_{p_k}}
-\frac{\partial A_{ij}}{\partial_{p_k}}\frac{\partial B_{i'j'}}{\partial_{x_k}}\Big )
\;\;, \end{equation} 
since, for example, 
$A\equiv A(x_k,p_k;X_i,P_i)=\sum_{i,j}A_{ij}(x_k,p_k)(X_i-iP_i)(X_j+iP_j)$. 

Thus, evolution of hybrid observables, of the density $\rho$ 
in particular, can induce a {\it structural change}: while continuing to be CL 
observables, they do not remain QM observables (quadratic forms in $X_i,P_i$). 

\subsection{Enlarged ``classical$\;\times\;$almost-classical algebra'' of hybrid observables} 

In order to maintain formal consistency of the algebraic framework, we assume: 
\\ \noindent $\bullet$ 
H) The algebra of hybrid 
observables is closed under the QM-CL Poisson bracket operation, implemented 
by $\{\;,\;\}_\times$ . 
\\ \noindent
This amounts to a {\it physical hypothesis} and its consequences will be discussed in the following. 
  
The normalization constraint, cf. Eq.\,(\ref{oscillnormalization}), 
is preserved under the evolution, since $\{{\cal C},{\cal H}_\Sigma\}_\times =0$, 
even in presence of QM-CL hybrid interaction. Consistently with closure of the 
enlarged algebra of hybrid observables, we also obtain:  
\begin{equation}\label{constraintG} 
\{{\cal C}(X_i,P_i),{\cal G}(x_k,p_k;X_i,P_i)\}_\times 
=\{{\cal C}(X_i,P_i),{\cal G}(x_k,p_k;X_i,P_i)\}_{\mbox{\scriptsize QM}} 
=0 
\;\;, \end{equation} 
where ${\cal G}(x_k,p_k;X_i,P_i)$ stands for any element of the 
enlarged algebra \cite{me11}. 

We define an {\it almost-classical observable} as a real-valued bilinear function of 
the phase space coordinates $(X_i,P_i)$ built from pairs of factors like 
$(X_i-iP_i)(X_j+iP_j)$, such as in the left-hand side of Eq.\,(\ref{nonlinear}). 
This implies: 
\\ \noindent $\bullet$ 
I) The QM observables (quadratic forms in phase space coordinates) 
form a subset of almost-classical observables which, in turn, form a subset of 
classical observables 
(real-valued regular functions of phase space coordinates).  
 
Furthermore, elements of the algebra of hybrid observables, 
generally, are {\it classical} with respect to 
coordinates $(x_k,p_k)$ and {\it almost-classical} with respect to coordinates $(X_i,P_i)$.  

The meaning of this enlarged 
``classical$\;\times\;$almost-classical algebra'' for interacting QM-CL hybrids 
has been illustrated in Ref.\,\cite{me12} by a {\it Gedankenexperiment}, which 
questions naive expectation that quantum and classical objects 
evolve separately in quantum and classical ways, when they no longer interact. -- 
The enlargement of the algebra of observables might be a hint that features 
of QM-CL hybrids could be relevant for 
{\it how QM emerges}. One would like to understand how a large  
algebra of classical observables (regular functions on phase space) is reduced, 
via almost-classical observables at an intermediary stage, to a smaller QM algebra 
(self-adjoint operators on Hilbert space) for an object that becomes ``quantized''.  

\subsection{Almost-classical observables in a toy model of an interacting QM-CL hybrid}

In order to further illuminate the necessity of an enlarged 
``classical$\;\times\;$almost-classical algebra'' of observables, 
as compared to the separable case in the absence of a genuine hybrid interaction, 
we present a simple model.  

Consider a CL object in one dimension together with a two-state QM object, 
described by phase space coordinates $(x,p)$ and $(X_\pm,P_\pm)$, respectively, in 
our formalism. We recall the Eqs.\,(\ref{rhoevol})--(\ref{HtotalInt}):  
$-\partial_t\rho = \{\rho ,{\cal H}_\Sigma\}_\times$, with the 
Hamiltonian function ${\cal H}_\Sigma:={\cal H}_{\mbox{\scriptsize CL}}(x,p)
+{\cal H}_{\mbox{\scriptsize QM}}(X_\pm,P_\pm)+{\cal I}(x,p;X_\pm,P_\pm)$ and with 
a particular 
hybrid interaction, ${\cal I}(p;X_\pm,P_\pm)\equiv\langle\psi |\hat{\cal I}(p)|\psi\rangle$, 
defined by: 
\begin{equation}\label{hybridI}
\hat{\cal I}(p):=g(t)p\hat\sigma_z
\;\;, \end{equation} 
where $\hat\sigma_z$ is a Pauli matrix and $g$ stands for a time-dependent coupling.   
  
This model can be easily solved, if we make a few additional simplifications in due course. --  
Define $f:=\int_0^T\mbox{d}tg(t)$, where $T$ represents the duration of the interaction. 
For sufficiently small $T$ and a strong coupling, we may neglect the influence of 
${\cal H}_{\mbox{\scriptsize CL}}+{\cal H}_{\mbox{\scriptsize QM}}$ on the evolution of the 
hybrid density $\rho$, in comparison with ${\cal I}$. In this case, the Liouville equation 
can be integrated with the result: 
\begin{equation}\label{rho1} 
\rho (x,p;X_\pm,P_\pm;T)=\rho (x-f\langle\psi |\hat\sigma_z|\psi\rangle,p;X_\pm,P_\pm;0)
\;\;, \end{equation} 
{\it i.e.}, in terms of the initial density.  
  
Let us assume that initially there were no correlations between CL and QM sectors.  
Therefore, the initial hybrid density is factorized, 
$\rho (x,p;X_\pm,P_\pm;0)=
\rho_{\mbox{\scriptsize CL}}(x,p;0)\rho_{\mbox{\scriptsize QM}}(X_\pm,P_\pm;0)\;$. 
Together with Eq.\,(\ref{rho1}), this implies: 
\begin{eqnarray} 
\rho (x,p;X_\pm,P_\pm;T)&=&
\rho_{\mbox{\scriptsize CL}}(x-f\langle\psi |\hat\sigma_z|\psi\rangle,p;0)
\rho_{\mbox{\scriptsize QM}}(X_\pm,P_\pm;0)
\nonumber \\ [1ex] \label{rho2}
&\equiv&
\rho_{\mbox{\scriptsize CL}}(x-f\langle\psi |\hat\sigma_z|\psi\rangle,p;0)
\langle\psi |\hat\rho_{\mbox{\scriptsize QM}}|\psi\rangle  
\;\;. \end{eqnarray} 

The state vector can always be expanded with respect to 
a twodimensional orthonormal basis $\{ |\pm\rangle\}$,  
$\sqrt 2|\psi\rangle =(X_++iP_+)|+\rangle +(X_-+iP_-)|-\rangle\;$. 
Furthermore, if the initial density 
matrix has a decohered form with respect to $\hat\sigma_z$, that is 
$\hat\rho_{\mbox{\scriptsize QM}}=w_+|+\rangle\langle +|+w_-|-\rangle\langle -|\;$,  
with $w_++w_-=1\;$, then 
$2\langle\psi |\hat\rho_{\mbox{\scriptsize QM}}|\psi\rangle 
=w_+(X_+^{\;2}+P_+^{\;2})+w_-(X_-^{\;2}+P_-^{\;2})$ and   
$2\langle\psi |\hat\sigma_z|\psi\rangle 
=(X_+^{\;2}+P_+^{\;2})-(X_-^{\;2}+P_-^{\;2})\;$. 

Recalling also the 
normalization condition, Eq.\,(\ref{oscillnormalization}), 
here $X_+^{\;2}+P_+^{\;2}+X_-^{\;2}+P_-^{\;2}=2\;$, 
we can interpret the result of Eq.\,(\ref{rho2}) in a simple way. -- 
Correlated with the probability to find a given $|\psi\rangle$ in the 
initial ensemble state of the QM sector, the position of the CL distribution 
is shifted by a certain amount to the left or right along the $x$-axis. 
In particular, if $|\psi\rangle$ has either $|+\rangle$- or $|-\rangle$-component only, 
then the shift amounts to $x\rightarrow x\mp f$, respectively. If, furthermore, the initial 
distribution is strongly peaked, such that  
$\rho_{\mbox{\scriptsize CL}}(x-f,p;0)\rho_{\mbox{\scriptsize CL}}(x+f,p;0)\approx 0$, 
then the CL degree of freedom acts like a ``pointer'' indicating the distribution 
of results of ``spin-up/down''-measurements effected on the QM two-state object: 
\begin{equation}\label{rho3}
\rho (x,p;X_\pm,P_\pm;T)=\rho_{\mbox{\scriptsize CL}}(x\mp f,p;0)\cdot w_\pm
\;\;, \end{equation} 
where either upper or lower signs apply. This describes an ideal measurement situation, 
where the {\it CL sector of the hybrid ``measures'' the QM sector}. 

We do not expect qualitative features of this model to change, if one or the other 
of many possible generalizations is incorporated. In particular, looking back at  
the quite general result in Eq.\,(\ref{rho1}), we see explicitly that $\rho$, as a result  
of the hybrid interaction chosen in Eq.\,(\ref{hybridI}), and for a generic initial 
$\rho_{\mbox{\scriptsize CL}}$, becomes unavoidably an element 
of the larger ``classical$\;\times\;$almost-classical algebra'' of hybrid observables   
defined in the first part of this Section~3.  
The way this happens is in accordance with our general discussion above.  

Similar effects must show up in the evolution of other hybrid observables $O$, 
determined by $\frac{\mbox{d}O}{\mbox{d}t}=\partial_tO+\{ O,{\cal H}_\Sigma \}_\times\;$, 
cf. Eq.\,(\ref{evolution}), in the presence of a genuine hybrid interaction.   

\section{Concluding remarks} 

We have presented a review of the linear dynamics of QM-CL hybrids laid out earlier 
in Refs.\,\cite{me11,me12}.  

We emphasize again that the CL sector of a hybrid does not necessarily present  
an approximation for some of the quantum mechanical degrees of freedom in a 
fully quantum mechanical multi-partite system. Rather, 
we have continued to study presently,  
whether such a hybrid theory can stand formally on 
its own and meet all the posed consistency requirements, cf. Section~1 .    

This has led us here to discuss in more detail the earlier 
observation that a consistent description of hybrids seems to entail an enlarged 
algebra of observables, in particular, as compared to the Cartesian product of sets of 
observables that belong to QM and CL sectors, in the separable case. 
This observation has been 
made more recently also in different context \cite{Buric12b,Barcelo12}.  

We recall that Man'ko and his collaborators repeatedly pointed out that classical 
states may differ from what could be obtained as the ``$\hbar\rightarrow 0$'' 
limit of quantum mechanical ones. Furthermore, they show that all states can 
be classified by their ``tomograms'' as {\it either} CL {\it or} QM, CL {\it and} QM, 
and {\it neither} CL {\it nor} QM~\cite{Manko04,Manko11}. 
Yet, in order to understand the origin of these ``Man'ko classes'', 
a dynamical explanation has been missing.

We find the results of Section~3. interesting in this respect, namely 
that a consistent hybrid description enforces the enlarged algebra of observables, 
due to genuine hybrid interactions. 
More specifically, we have seen in detail how QM observables have to be generalized 
in the form intermediary {\it almost-classical observables}, which form a subset of 
corresponding classical observables. This corresponds to  
Man'ko's findings in our approach, where all states are represented in phase space, 
and provides a dynamical explanation. 

In a simple model, we have shown that this enlargement of the set of observables 
is not only necessary but very wellcome. It is generated dynamically and accommodates 
the measurement-like effect of  the hybrid interactions. Through 
them, the would-be (in the separable case)  CL degrees of freedom of the hybrid 
perform measurements on the would-be QM ones.   

It would be most interesting, if one could similarly find 
some underlying dynamical reason for such {\it structural change} that occurs when, 
conversely, a CL object is turned into a QM object, {\it i.e.}, when it {\it becomes} quantized. 

\section*{Acknowledgments}

It is a great pleasure to thank Marco Genovese and his co-organizers for the invitation 
to the 6th Workshop ad memoriam of Carlo Novero {\it ``Advances in Foundations 
of Quantum Mechanics and Quantum Information with Atoms and Photons''} (Torino, May 2012), 
where part of this work was presented, and to thank Vladimir Man'ko, Christof Wetterich, 
Nikola Buri\'c, and Marcel Reginatto for discussions on various occasions.   


\end{document}